# The structures and thermoelectric properties of the infinitely adaptive series $(Bi_2)_m(Bi_2Te_3)_n$

J.W.G. Bos[a], H.W. Zandbergen[b], M.-H. Lee[c], N.P. Ong[c] and R. J. Cava[a]

[a] Department of Chemistry, Princeton University, Princeton, New Jersey 08544, USA

[b] National Centre for High Resolution Electron Microscopy, Department of Nanoscience, Delft Institute of Technology, Delft, The Netherlands.

[c] Department of Physics, Princeton University, Princeton, New Jersey 08544, USA

The structures and thermoelectric properties of the $(Bi_2)_m(Bi_2Te_3)_n$ homologous series, derived from stacking hexagonal $Bi_2$ and $Bi_2Te_3$ blocks, are reported. The end-members of this series are metallic Bi and semiconducting $Bi_2Te_3$; nine members of the series have been studied. The structures form an infinitely adaptive series and a unified structural description based on a modulated structure approach is presented. The as-synthesized samples have thermopowers (S) that vary from *n*-type for $Bi_2Te_3$ to *p*-type for phases rich in $Bi_2$ blocks but with some $Bi_2Te_3$ blocks present, to *n*-type again for Bi metal. The thermoelectric power factor ($S^2/\rho$) is highest for Bi metal (43 $\mu W/K^2$ cm at 130 K), followed by $Bi_2Te_3$ (20 $\mu W/K^2$ cm at 270 K), while $Bi_2Te$ (m:n = 5:2) and $Bi_7Te_3$ (m:n = 15:6) have 9 $\mu W/K^2$ cm (at 240 K) and 11 $\mu W/K^2$ (at 270 K), respectively. The results of doping studies with Sb and Se into $Bi_2Te$ are reported.

**Introduction**

Materials suitable for thermoelectric refrigeration and power generation have attracted renewed interest in the past decade. A large number of new materials with promising thermoelectric properties have been investigated. These include, for example, filled skutterudites,[1, 2] clathrates,[3] pentatellurides,[4] half-heusler alloys,[5-7] ternary and quaternary heavy metal chalcogenides,[8-11] and layered cobalt oxides.[12-14] In addition, the efficiency of well known thermoelectric materials such as $Bi_2Te_3$ and PbTe has been improved dramatically by incorporation into nano-structured devices.[15, 16]

Thermoelectric materials that function well at or below room temperature are particularly desirable. Among the most suitable materials are the long-known $Bi_2Te_3$ and Bi based alloys. Given the importance of these alloys, we have investigated the structures and thermoelectric properties of the $(Bi_2)_m(Bi_2Te_3)_n$ homologous series. This series spans the phase-space from Bi to $Bi_2Te_3$ with intermediate compositions such as BiTe. These phases, which have not attracted much interest, were obtained by low temperature synthesis. The crystal structures are derived from an ordered stacking of $Bi_2Te_3$ and $Bi_2$ building blocks (illustrated in Fig. 1). All phases in the bismuth-tellurium phase diagram excluding elemental Te are members of the homologous series. The known members are $Bi_2Te_3$ (m:n = 0:3), $Bi_4Te_5$ (m:n = 1:5), $Bi_6Te_7$ (m:n = 2:7), $Bi_8Te_9$ (m:n = 3:9), BiTe (m:n = 1:2), $Bi_4Te_3$ (m:n = 3:3), $Bi_2Te$ (m:n = 2:1), $Bi_7Te_3$ (m:n = 15:6), and Bi (3:0), while the existence of others, such as $Bi_{10}Te_9$ (m:n = 6:9), has been predicted.[17]

Band structure calculations on compounds in the same structural family in the Bi-Se system suggest that the $Bi_2$ double layers and $Bi_2Se_3$ units each have a formal valence of zero.[18, 19] The adding of zerovalent Bi double layers therefore is not expected to change the valence of pristine $Bi_2Se_3$ layers and therefore may explain the formation of the series between what appears to be a simple ionic compound ($Bi_2Se_3$) and a metal (Bi).[18] These calculations also showed that the energy of formation of the $(Bi_2)_m(Bi_2Se_3)_n$ phases is close to zero.[19] This suggests that any ratio of m:n might exist in a crystallographically ordered series of distinct phases made from stacking $Bi_2Se_3$ and $Bi_2$ building blocks. In the theoretical picture, the interaction between neighboring $Bi_2Se_3$ blocks is of the Van der Waals type, whereas interactions involving $Bi_2$ blocks are of weak covalent nature. The calculations show that $Bi_2Se_3$ should behave as a narrow gap semiconductor, while the addition of $Bi_2$ blocks results in band structures characteristic of a semimetal.

Investigations into the thermoelectric properties of the $(Bi_2)_m(Bi_2Te_3)_n$ homologous series have mainly focused on $Bi_2Te_3$ and Bi.[20, 21] The intermediate phases have either not been investigated in detail or have not been recognized as distinct thermoelectric phases. No correlation of structure and properties of the sort we report here has been previously established. An example of previous work can be found in the 1995 CRC handbook of thermoelectrics.[22] In that study, "$Bi_2Te_3$ single crystals" with compositions between 50-70 atomic percent tellurium were studied. These crystals were grown by a traveling heater method at temperatures between 560-580 $^0$C. From our work on polycrystalline samples we know that annealing in this temperature range (for compositions 30-60 at% Te) results in a two-phase mixture of Bi and $Bi_2Te_3$, and that the $(Bi_2)_m(Bi_2Te_3)_n$ homologous series is only stable at lower temperatures. The low temperature growth of $Bi_{1+x}Te_{1-x}$ thin films (substrate temperature 250 $^0$C) did result in the stabilization of several of the members of the $(Bi_2)_m(Bi_2Te_3)_n$ homologous series.[23] It was found in that work that the Seebeck coefficient changed from negative to positive between 0.56 and 0.59 atomic fraction Bi. The structures and thermoelectric properties of $Sb_2Te$ (m:n = 2:1), $Sb_4Te_3$ (m:n = 6:6), $SbTe$ (m:n = 1:2), and $Sb_8Te_9$ (m:n = 3:9), members of the $(Sb_2)_m(Sb_2Te_3)_n$ homologous series, were recently reported.[24] The Seebeck coefficients at room temperature are between 10-40 µV/K.

In this study we correlate the crystal structures and thermoelectric properties of the $(Bi_2)_m(Bi_2Te_3)_n$ homologous series. (A list of the studied compositions is given in Table 1.) The thermopower varies systematically from *n*-type for $Bi_2Te_3$ to *p*-type for phases rich in $Bi_2$ blocks but with some $Bi_2Te_3$ blocks present, to *n*-type again for Bi metal. The most promising power factors (PF = $S^2/\rho$) are 43 µW/K$^2$ cm at 130 K for Bi, 20 µW/K$^2$ cm at 270 K for $Bi_2Te_3$, 9 µW/K$^2$ cm at 240 K for $Bi_2Te$ (m:n = 5:2) and 11 µW/K$^2$ cm at 270 K for $Bi_7Te_3$ (m:n = 15:6). The results of Sb and Se doping studies into $Bi_2Te$ are reported: the highest power factor was found for $Bi_2Te_{0.67}Se_{0.33}$ (10 µW/K$^2$ cm at 190 K).

**Experimental**

Polycrystalline samples of several members of the $(Bi_2)_m \cdot (Bi_2Te_3)_n$ homologous series were prepared by standard solid state reaction. All samples were prepared by elongated heating at low temperatures. Stoichiometric amounts of pulverized Bi (99.99%) and Te (99.99%) pieces were mixed together and vacuum sealed in quartz tubes. After an initial heating step of 1 day, the samples were homogenized using mortar and pestle and pressed into pellets. These were subsequently heated for 2 weeks with one intermediate regrinding. The following temperatures were used: $Bi_2Te_3$ (m:n = 0:3) at 525 $^0$C, $Bi_4Te_5$ (m:n = 1:5), $Bi_6Te_7$ (m:n = 2:7), $Bi_8Te_9$ (m:n = 3:9) and BiTe (m:n = 1:2) at 485 $^0$C,

Bi$_4$Te$_3$ (m:n = 1:1) at 375 $^0$C, Bi$_2$Te (m:n = 2:1) and Bi$_7$Te$_3$ (m:n = 15:6) at 285 $^0$C, and Bi (m:n = 3:0) at 255 $^0$C. Attempts to prepare compositions with m:n > 15:6 resulted in mixed phase samples of Bi and Bi$_7$Te$_3$. Samples with m:n < 1:5 were not investigated. The Bi$_{2-x}$Sb$_x$Te and Bi$_{2-x}$Sb$_x$Te$_{0.67}$Se$_{0.33}$ (0 ≤ x ≤ 0.3) compounds were also prepared following the procedure described above with a synthesis temperature of 285 $^0$C. Higher values of x were not tried. The phase purity of all samples was confirmed by X-ray powder diffraction (XRPD) using a Bruker D8 Focus diffractometer with Cu K$_\alpha$ radiation fitted with a scintillation counter and diffracted beam monochromator. Data were collected between 5 ≤ 2θ ≤ 100$^0$ with a 0.02$^0$ step size and counting times of 5s/step. The lattice constants and modulation vectors for the prepared (Bi$_2$)$_m$(Bi$_2$Te$_3$)$_n$ phases were obtained from a LeBail fit of the XRPD patterns using the JANA2000 program.

Electron microscopy analysis was performed with Philips CM300UT electron microscopes having a field emission gun and operated at 300 kV. Electron transparent areas of specimens were obtained by crushing them slightly under ethanol to form a suspension and then dripping a droplet of this suspension on a carbon-coated holey film on a Cu grid.

Thermopower data (S = V/ΔT) were collected using an MMR technologies SB100 Seebeck measurement system. Rectangular bars (app. 1 × 1 × 5 mm) were mounted on a sample holder using silver paint. Electrical resistivities were measured using the resistance bridge in a Quantum Design Physical Properties Measurements System (PPMS). Contacts were made using platinum wire and silver paint in a standard four-point geometry.

**Structure**

The commonly used structural description of the $(Bi_2)_m(Bi_2Te_3)_n$ homologous series is based on an ordered stacking of two layer Bi-Bi and five layer Te-Bi-Te-Bi-Te blocks.[17] Using this description, possible structures of the studied binaries can be predicted, and are illustrated in Fig. 1. Typical of infinitely adaptive structures[25], more than one stacking sequence can be envisioned for a given m:n ratio, though the compounds themselves pick one structure. The Bi and Te layers that make up the building blocks are ABC stacked and the total number of layers per unit cell is a multiple of three. For this reason, both trigonal and rhombohedral structures occur, with a rhombohedral structure if (n+m)/2 is even (Fig. 1). The observed *a*-parameters (the dimensions of the layers in-plane) are very similar (around 4.4 Å), while the *c*-parameters (the repeat along the stacking direction) can be predicted from:

$$c_{predicted} = \frac{1}{3}[mc'+nc''] \qquad (1)$$

where *c′* and *c′′* are the *c*-parameters of the end members Bi and $Bi_2Te_3$ respectively, and m and n are the numbers of $Bi_2$ and $Bi_2Te_3$ blocks per unit cell. The predicted *c*-parameters are given in Table 1. They vary widely and irregularly with m:n ratio (Bi-fraction). This can also be appreciated from Fig. 1. Where experimental data is available, the predicted and experimental *c*-parameters are in good agreement (Table 1).

The irregular dependence of the *c*-parameter on m:n ratio appears to be in contradiction with the XRPD patterns, shown in Fig. 2, which suggest a smooth variation of the lattice constants in the series. This behavior has also been observed for the $(Bi_2)_m(Bi_2Se_3)_n$ homologous series,[26] and suggested that the structures can be best described in terms of the structural modulation of an average structure. This approach to the series was also followed in the current study. We successfully indexed the electron diffraction (Fig. 3) and XRPD patterns for the $(Bi_2)_n(Bi_2Te_3)_m$ series using a basic subcell ($a \sim 4.4$ Å, $c \sim 6.0$ Å) and a modulation of the structure along the c-axis.

Electron diffraction studies (a representative example is shown in Fig. 3) showed strong reflections that could be indexed on a basic hexagonal subcell, and weaker reflections, due to the structural modulation, that were indexed by introducing a modulation vector $\mathbf{q} = \gamma[001]^*$. This approach has been used to describe other structures of the "infinitely adaptive" type,[25] as described in more detail below. Quantitative determination of the subcell lattice constants and the $\gamma$ values were obtained from a 4 dimensional LeBail fit to the XRPD patterns. The obtained lattice constants, $\gamma$ values, and goodness of fit parameters are listed in Table 1. Note that the structural analysis presented here is limited to the cell constants and modulation vectors. The solution of the exact 4-dimensional structure is beyond the scope of the present study and will be presented elsewhere.

The lattice constants and $\gamma$ values are shown in Fig. 4 for different members of the homologous series, as a function of the Bi fraction. The subcell constants show two linear regimes, with the boundary at a Bi fraction of 0.47, which corresponds to $Bi_8Te_9$. The $\gamma$ value, in contrast, has a linear, continuous dependence on the Bi fraction. The $\gamma$ values for the following compositions yield rational superstructures: $Bi_2Te_3$ ($\gamma = 6/5$), $Bi_4Te_3$ ($\gamma = 9/7$), $Bi_2Te$ ($\gamma = 4/3$) and Bi ($\gamma = 3/2$). These compositions can therefore be described using a standard 3-dimensional superstructure, if desired. The calculated conventional superstructure c-axes are in good agreement with both the predicted and literature values (Table 1). The crystal structure of $Bi_2Te$ can be described, for example, either with a conventional unit cell of a ~ 4.465 Å and c ~ 17.915 Å containing 2 $Bi_2$ blocks and 1 $Bi_2Te_3$ block, or as a structural modulation of a subcell with a ~ 4.465 Å, and c ~ 5.972 Å, with a γ value of 4/3; the conventional description allows a simple intuitive picture of the individual structure to be envisioned, but the modulated structure description allows this compound to be put into the context of the description of the whole series, which includes both commensurate and incommensurate superstructures.

The γ values for the other compositions are not rational, corresponding to incommensurate modulations, and their structures can only be approximated in a 3-dimensional picture. The XRPD patterns cannot be indexed with the predicted *c*-parameters for commensurate superlattices. Note that for most compositions in Table 1, commensurate superstructures have been previously reported; the present study shows many of them in fact to have incommensurate structural modulations.

The description of the Bi-Te phases in terms of this structural scheme provides a uniform framework for describing all the structures in the Bi-Te system between $Bi_2Te_3$ and Bi, and yields a good framework for describing the relation between composition, cell constants and γ, that is more informative, for example, then describing this important region in the Bi-Te phase diagram in terms of "% Bi", as is frequently done. Our structural characterization of the $(Bi_2)_n(Bi_2Te_3)_m$ series shows it to be a classical example of an "infinitely adaptive series", as originally proposed by Anderson[25]. In such a series, changes in chemical composition result in neither a progression of phases separated by two-phase regions, nor disordered nonstoichiometric solid solutions. In an infinitely adaptive series, made by stacking building blocks of fixed composition (in the current case $Bi_2$ and $Bi_2Te_3$ blocks) in different ratios, infinitesimal changes in composition result in distinct, fully ordered structures, sometimes with very long periodicity. In the cases where the composition corresponds to ratios of building blocks that are small integers, conventional crystal structure determinations can be performed, and 3D structures can be envisioned (e.g. $Bi_2Te_3$ and $Bi_2Te$ in the current case), but all members of the series, with both commensurate and incommensurate structural modulations, are conceptually the same.

The diffraction patterns for the structures in an infinitely adaptive series yield important information on their character. Strong subcell reflections yield the average structure: in the case of the current series the average in-plane size and the average height of the simple stacking layer thickness. The ideal average, substructure dimensions can be calculated from the basic unit cell parameters of the end members, in our case $Bi_2Te_3$ and Bi and from the fraction of layers contained in the $Bi_2$ and $Bi_2Te_3$ blocks:

$$a(n,m) = \left(\frac{2m}{2m+5n}\right) a_{Bi-basic} + \left(\frac{5n}{2m+5n}\right) a_{Bi_2Te_3-basic} \qquad (2)$$

$$c(n,m) = \left(\frac{2m}{2m+5n}\right) c_{Bi-basic} + \left(\frac{5n}{2m+5n}\right) c_{Bi_2Te_3-basic} \qquad (3)$$

These calculated parameters are shown in figure 4 as solid lines (calculated points indicated). The trends can be understood as follows: adding $Bi_2$ blocks to $Bi_2Te_3$ results in a strain at the interface ($a_{Bi} > a_{Bi2Te3}$) and as a result the *a*-axis expands. The *c*-axis on the other hand contracts on adding $Bi_2$ blocks as $c_{Bi} < c_{Bi2Te3}$.

Although it can be seen that the general behavior is as expected, the observed average cell parameters deviate substantially from the ideal values determined by simple stacking of independent layers. As Anderson pointed out,[25] an infinitely adaptive series is in fact only expected to form if interactions between the building blocks are significant, with at least one of the structural components undergoing a structural distortion in the process of forming the phases. The effects of this kind of interaction are very likely manifested in the deviation of our observed average structure dimensions from the ideal values. A good illustration of this can be seen in figure 5, where we have plotted the observed c/a ratio (upper panel) and observed unit cell volume (lower panel) compared to the ideal values derived from stacking independent layers. The figure shows that although the shape of the layers deviates substantially from the ideal values across the composition range of the series, the cell volume behaves in a nearly ideal fashion. In other words, the molar volume behaves as expected though the shape of the layers is anomalous across the series. This can most straightforwardly be interpreted as being the result of exactly the kind of strain induced structural distortion needed to hold the series together.

The clear change in character of the structure at a Bi fraction of 0.47 is intriguing. It could be that at this particular ratio of $Bi_2$ and $Bi_2Te_3$ blocks the type of structural distortion that occurs to accommodate the layers changes in character, i.e. that the change is fully structurally driven. Alternatively, the change could be electronically driven, Although band structure calculations have suggested that the energy of formation of the homologous series is close to zero and that the $Bi_2$ and $Bi_2Te_3$ blocks can be regarded as neutral species, the changes in the slope of c/a observed at a Bi fraction of 0.47 could reflect a change in the electronic interactions between the blocks at that composition rather than a difference in strain mechanism. The location of the crossover does not correlate with an obvious chemical border, however, such as the change from Te to Bi rich compositions. More detailed theoretical consideration of this structural series from a structural and electronic perspective would be of great interest.

**Thermoelectric Properties**

The electrical resistivity and thermopower measurements of the $(Bi_2)_m(Bi_2Te_3)_n$ homologous series are summarized in Figs. 6 and 7. From the resistivity measurements it is clear that $Bi_2Te_3$ (m:n = 0:3) and $Bi_4Te_3$ (m:n = 3:3) are somewhat anomalous. Starting from $Bi_2Te_3$, the introduction of $Bi_2$ blocks leads to much weaker temperature dependencies and substantially lower resistivities (over most of the temperature range) for the compositions between $Bi_4Te_5$ (m:n = 1:5) and BiTe (m:n = 1:2). Adding more $Bi_2$ blocks leads to $Bi_4Te_3$ (m:n = 3:3), which has a temperature dependence characteristic of a compensated semiconductor. Adding yet more $Bi_2$ blocks results in a decrease in resistivity and a return to a temperature dependence characteristic of a poor metal. The lowest resistivities are observed for Bi-metal and the compositions rich in $Bi_2Te_3$ blocks.

The thermopower measurements (fig. 7) reveal a change from *n*- to *p*-type with an increasing fraction of $Bi_2$ blocks. Starting again from $Bi_2Te_3$, *n*-type behavior is observed, with a maximum value of -175 µV/K at 400 K. The compositions between $Bi_4Te_5$ (m:n = 1:5) and BiTe (m:n = 1:2) are all *n*-type and have similar magnitudes and temperature dependencies of the Seebeck coefficient. Note that these correspond to the compositions with the lowest resistivities. For $Bi_4Te_3$ (m:n = 3:3), a crossover between *p*- and *n*-type is observed at 175 K, as described above, and this composition also defines the change from *n*- to *p* type conduction in the homologous series. The higher $Bi_2$ block containing samples are *p*-type. A maximum thermopower of +90 µV/K is observed for $Bi_2Te$ (m:n =

2:1). Bi-metal on the other hand, prepared in the same fashion as the rest of the series is *n*-type with a maximum thermopower of -70 µV/K.

The temperature dependence of the thermoelectric power factors (PF = $S^2/\rho$) are shown in Fig. 8 for representative members of the $(Bi_2)_m(Bi_2Te_3)_n$ homologous series. Bi has the highest power factor at low temperatures (43 µW/K$^2$ cm at 130 K), reflecting its low resistivity and medium sized thermopower. $Bi_2Te_3$ has $PF_{max}$ = 20 µW/K$^2$ cm at 270 K. From the intermediate compositions, the most promising are $Bi_2Te$ (m:n = 2:1) and $Bi_7Te_3$ (m:n = 15:6) that have 9 µW/K$^2$ cm (at 240 K) and 11 µW/K$^2$ (at 270 K), respectively.

An overview of the important structural and transport properties of the homologous series is given in Fig. 9. The *c/a*-axis ratio reveals the presence of two linear regimes with the crossover occurring near a Bi fraction of 0.47, as described previously. The residual resistivity ratio (RRR), defined by the ratio of the resistivity at 290 K and that at 5 K, is shown. $Bi_2Te_3$ has RRR = 22 and has the temperature dependence of a good metal. All other compositions have RRR < 5, and have temperature dependences of their resistivities more characteristic of degenerate semiconductors or poor metals. This includes the polycrystalline Bi sample, which has RRR = 5. Note that compositions that can be described using a conventional commensurate three dimensional crystal structure do not have significantly different RRR values. This suggests that there is no fundamental electronic difference between compositions with commensurate or incommensurate modulations. It is clear from our data that the addition of only a few $Bi_2$ blocks to $Bi_2Te_3$ results in a profound change in the electrical conduction. Insertion of only a few $Bi_2$ blocks in $Bi_2Te_3$ results in a 5-6 fold reduction of the Seebeck coefficient at 290 K. The Seebeck coefficient then changes sign around a Bi fraction of 0.57 ($Bi_4Te_3$). For compositions rich in $Bi_2$ blocks the Seebeck coefficient is ~ +90 µV/K, while it is negative again for Bi. As mentioned above, the power factor of Bi metal is the highest, followed by that of $Bi_2Te_3$. Of the intermediate members $Bi_2Te$ (m:n = 2:1) and $Bi_7Te_3$ (m:n = 5:2) have the most potential.

The relatively high thermopower of $Bi_2Te$ made it interesting for further study. In the remainder of this section the $Bi_{2-x}Sb_xTe$ and $Bi_{2-x}Sb_xTe_{0.67}Se_{0.33}$ (0 ≤ x ≤ 0.3) substitution series are described. Both $Bi_{2-x}Sb_xTe$ and $Bi_{2-x}Sb_xTe_{0.67}Se_{0.33}$ were found to be phase pure from XRPD, and their patterns could be indexed using the superspace formalism described above. The cell constants and γ values are given in Fig. 10. The *a, c*-axes and cell volume for $Bi_{2-x}Sb_xTe$ decrease (almost) linearly up to x = 0.2 and after that a sudden drop is observed. The γ parameter decreases linearly in the entire doping range. The $Bi_{2-x}Sb_xTe_{0.67}Se_{0.33}$ series also shows a decrease in cell constants albeit much less regularly. In spite of the different x-dependence, the magnitude of the observed reductions in cell constants between x = 0 and x = 0.3 are similar for both series. The γ parameter for $Bi_{2-x}Sb_xTe_{0.67}Se_{0.33}$ decreases linearly between 0.1 ≤ x ≤ 0.3 while for x = 0 an anomalously small value is found. This suggests that $Bi_2Te_{0.67}Se_{0.33}$ is somewhat different from the other members of the series. A view corroborated by the physical measurements below.

For $Bi_{2-x}Sb_xTe$, the room temperature resistivity decreases from ~ 1 mΩ cm (x = 0) to ~ 0.7 mΩ cm for x = 0.2 and x = 0.3 (Fig. 11a). The thermopower (Fig. 9b) decreases from around 90 µV/K (x = 0) to 50 µV/K (x = 0.3) at the same time, however.

The maximum values for the power factors are given in the inset in Fig. 11a. For x = 0.2, $PF_{max}$ = 10 µW/K$^2$ cm at 240 K, moderately higher than that for x = 0.

The electrical resistivity for Bi$_2$Te$_{0.67}$Se$_{0.33}$ (Fig. 12a) is similar in magnitude and temperature dependence to that of Bi$_2$Te. Its thermopower (Fig. 12b), however, peaks at a lower temperature (200-225 K). This results in a $PF_{max}$ of 10 µW/K$^2$ cm at 190 K (inset in Fig. 12a), compared to an optimal 240 K for Bi$_2$Te. The resistivities of the Bi$_{2-x}$Sb$_x$Te$_{0.67}$Se$_{0.33}$ samples have a slightly weaker dependence on temperature. For x = 0.1 and x = 0.2 the room temperature resistivities are slightly smaller, while for x = 0.3 the resistivity is slightly larger. The maximum thermopower decreases from 90 µV/K to 40-50 µV/K and moves to higher temperatures (~250K), similar to those observed for the Bi$_{2-x}$Sb$_x$Te series. This results in a rapid drop in thermoelectric power factor with doping in this group, with maximum values below 4 µW/K$^2$ cm (inset in Fig. 12a).

**Conclusions**

Using low temperature synthesis, we have been able to synthesize nine members of the infinitively adaptive (Bi$_2$)$_m$(Bi$_2$Te$_3$)$_n$ series. These materials consist of a stacked arrangement of Bi$_2$ and Bi$_2$Te$_3$ blocks. Electron diffraction and X-ray powder diffraction patterns were indexed using a superspace formalism. This resulted in a unified structural description using a basic hexagonal cell and modulation vector q = γ[001]$^*$. The γ values vary linearly with Bi-fraction, while for the *a*- and *c*-axis two linear regions are found. The latter reveal a volume conserving distortion of the Bi$_2$ and Bi$_2$Te$_3$ blocks that facilitates the formation of the infinitely adaptive series. The γ values for Bi$_2$Te$_3$, Bi$_4$Te$_3$, Bi$_2$Te and Bi were found to be rational, and these structures can be adequately described using the supercells shown in Fig. 1. The other prepared compositions cannot be described as in Fig. 1, but do belong to the $(Bi_2)_m(Bi_2Te_3)_n$ series and are in no way different, a view supported by our modulated structure analysis and physical property measurements.

Thermopower measurements revealed a change from *n*-type behavior for $Bi_2Te_3$ (-175 µV/K) to *p*-type behavior for phases rich in $Bi_2$ blocks but with some $Bi_2Te_3$ blocks present (+90 µV/K), to *n*-type again for Bi metal (-70 µV/K). The potential for application in thermoelectric applications was estimated from the thermoelectric power factor (PF = $S^2/\rho$). The PF is highest for Bi metal (43 µW/K$^2$ cm at 130 K), followed by $Bi_2Te_3$ (20 µW/K$^2$ cm at 270 K). From the intermediate compositions, $Bi_2Te$ (m:n = 5:2) and $Bi_7Te_3$ (m:n = 15:6) are the most promising with PF's of 9 µW/K$^2$ cm (at 240 K) and 11 µW/K$^2$ (at 270 K), respectively. The thermoelectric properties of $Bi_2Te$ were improved moderately by substitution of Sb on the Bi positions. Substitution of Se on the Te sites led to $Bi_2Te_{0.67}Se_{0.33}$, which is built from $Bi_2$ and $Bi_2Te_2Se$ blocks. The latter material has a maximum power factor of 10 µW/K$^2$ cm at 190 K (compared to 240 K for $Bi_2Te$).

**Acknowledgements**

This work was supported by the Air Force Research Laboratory. JWB acknowledges the Royal Society of Edinburgh for a Scottish Executive Personal Research Fellowship.

**Figure captions**

**Fig. 1.** Schematic crystal structures of compounds in the $(Bi_2)_m(Bi_2Te_3)_n$ homologous series. $Bi_2$ and $Bi_2Te_3$ blocks are represented by narrow white and broad gray rectangles. The atom positions within the blocks are shown at the end member compositions $Bi_2Te_3$ and Bi. Stacking sequences based on the ordered arrangement of $Bi_2$ and $Bi_2Te_3$ blocks are shown.

**Fig. 2.** X-ray powder diffraction patterns for the $(Bi_2)_m(Bi_2Te_3)_n$ homologous series. Selected reflections are indexed using four indices (hklm) to specify the basic cell and modulation quantities.

**Fig. 3.** Electron diffraction pattern for $Bi_4Te_5$. The block stacking direction (the *c*-axis) is vertical. Diffraction spots corresponding to the basic cell are marked by open squares, and the reflections indexed by the structural modulation vector $\mathbf{q} = \gamma[001]^*$ are marked by arrows.

**Fig. 4.** (a) The *a*-axis of the basic cell. (b) The *c*-axis of the basic cell. The dashed lines are guides to the eye. (c) Dependence of the γ values, defining the structural modulation $\mathbf{q} = \gamma[001]^*$, on Bi fraction. Triangles / solid lines correspond to ideal behavior (see text)

**Fig. 5.** Upper panel: The composition dependence of the observed c/a ratio in the $(Bi_2)_m(Bi_2Te_3)_n$ homologous series compared to the ideal ratio (solid line / triangles) calculated by using the *a* and *c* values of the end-members (see text). Lower panel, the composition dependence of the observed subcell volume compared to the calculated volume. Inset, the deviation of the volume from the calculated value.

**Fig. 6.** Temperature dependent electrical resistivities for the $(Bi_2)_m(Bi_2Te_3)_n$ homologous series.

**Fig. 7.** Temperature dependent thermopower for the $(Bi_2)_m(Bi_2Te_3)_n$ homologous series.

**Fig. 8.** Thermoelectric power factors (PF = $S^2/\rho$) for selected $(Bi_2)_m(Bi_2Te_3)_n$ compositions.

**Fig. 9.** The composition dependence of the c/a ratio (a), the residual resistivity ratio (b), the 290 K thermopower (c), and the 290 K power factors (d) for the $(Bi_2)_m(Bi_2Te_3)_n$ homologous series.

**Fig. 10.** The composition dependence of the lattice constants, unit cell volume and γ for $Bi_{2-x}Sb_xTe$ and $Bi_{2-x}Sb_xTe_{0.67}Se_{0.33}$. The dashed and dotted lines are guides to the eye. The solid lines correspond to linear fits.

**Fig. 11.** Electrical resistivities and thermopowers for $Bi_{2-x}Sb_xTe$. The inset in (a) shows the x-dependence of the maximum power factors.

**Fig. 12.** Electrical resistivities and thermopowers for $Bi_{2-x}Sb_xTe_{0.67}Se_{0.33}$. The inset in (a) shows the x-dependence of the maximum power factors.

Table 1. Refined lattice constants, unit cell volumes, modulation vectors, goodness of fit parameters, supercell *c*-axes, predicted supercell axes and ICSD *c*-parameters for $(Bi_2)_m (Bi_2Te_3)_n$. All fits were done using superspace group P: R-3m : -11.

| (m:n)[a] | formula | Bi fraction | $a$ (Å) | $c$ (Å) | Vol (Å$^3$) | $\gamma[001]^*$ | $\chi^2$ (-) | $R_{wp}$ (%) | $c_{supercell}$ (Å)[b] | $c_{predicted}$ (Å)[c] | ICSD and Ref. 17 c (Å) |
|---|---|---|---|---|---|---|---|---|---|---|---|
| 0:3 | $Bi_2Te_3$ | 0.40 | 4.3824(1) | 6.0947(1) | 101.369(4) | 1.2001(1) | 1.20 | 11.5 | 30.474 | 30.474 | 30.44-30.50 |
| 1:5 | $Bi_4Te_5$ | 0.44 | 4.4154(1) | 6.0284(2) | 101.782(6) | 1.2291(1) | 1.30 | 11.9 | - | 54.742 | 54.33 |
| 2:7 | $Bi_6Te_7$ | 0.46 | 4.4240(1) | 6.0169(2) | 101.984(6) | 1.2385(1) | 1.31 | 13.1 | - | 79.011 | 78.20 |
| 3:9 | $Bi_8Te_9$ | 0.47 | 4.4254(2) | 6.0077(2) | 101.893(7) | 1.2434(1) | 1.42 | 13.9 | - | 103.279 | 103.90 |
| 1:2 | BiTe | 0.50 | 4.4334(1) | 6.0005(1) | 102.139(4) | 1.2563(1) | 1.29 | 12.2 | - | 24.269 | 24.00 |
| 3:3 | $Bi_4Te_3$ | 0.57 | 4.4440(1) | 5.9887(2) | 102.426(6) | 1.2860(1) | 1.25 | 11.2 | 41.922 | 42.332 | 41.89 |
| 2:1 | $Bi_2Te$ | 0.67 | 4.4652(2) | 5.9718(3) | 103.114(7) | 1.3336(1) | 1.20 | 11.3 | 17.915 | 18.064 | - |
| 15:6 | $Bi_7Te_3$ | 0.70 | 4.4721(2) | 5.9707(3) | 103.414(7) | 1.3424(2) | 1.52 | 12.6 | - | 120.241 | 119.0 |
| 3:0 | Bi | 1.00 | 4.5452(1) | 5.9294(1) | 106.083(4) | 1.50 | 1.35 | 11.9 | 11.859 | 11.859 | 11.862 |

[a] (m/n) refers to the number of Bi and $Bi_2Te_3$ blocks per unit cell.
[b] When $\gamma$ is a rational number the superstructure is commensurate.
[c] Calculated from c = 1/3 (mc' + nc"), c' = 11.589 Å, c" = 30.474 Å.

Fig. 1

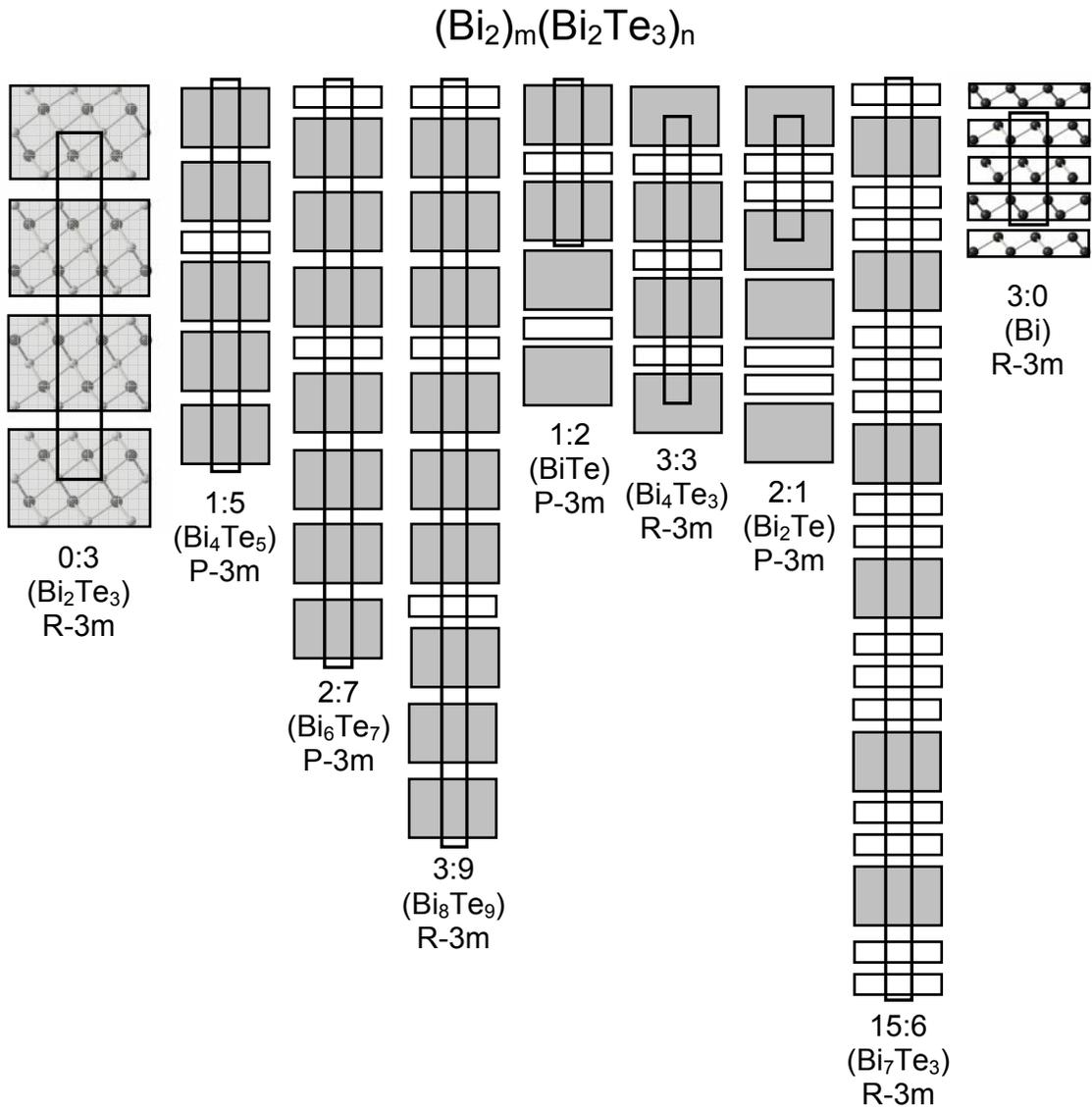

Fig. 2

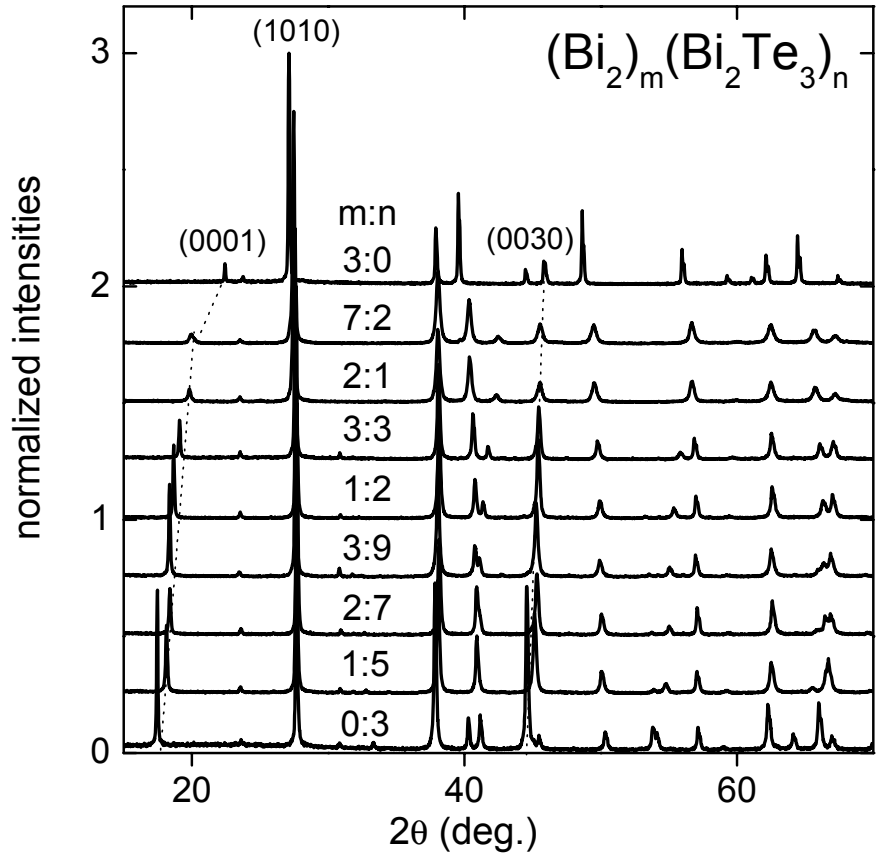

Fig. 3

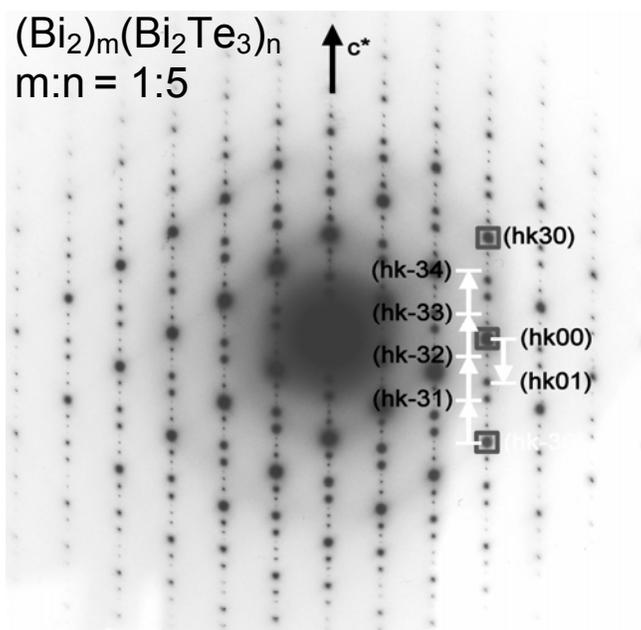

Fig. 4a

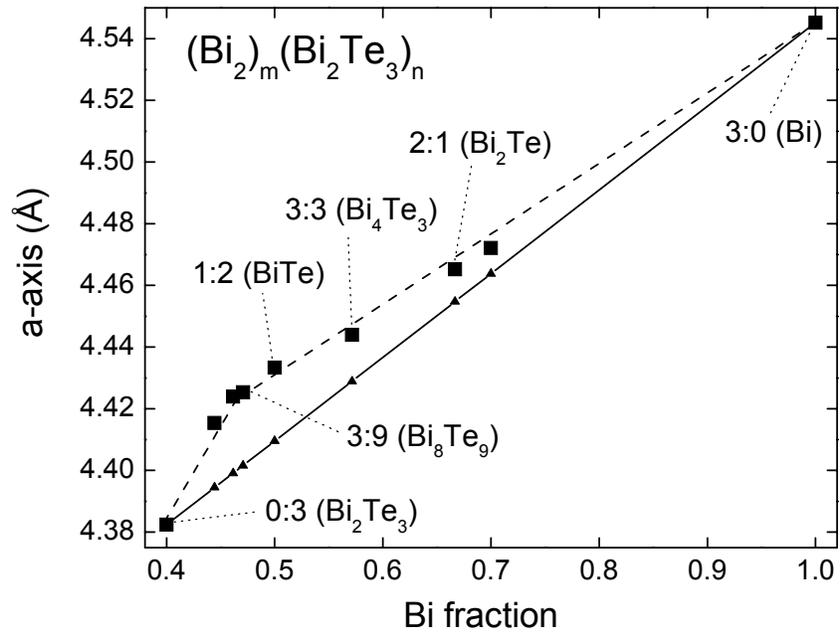

Fig. 4b

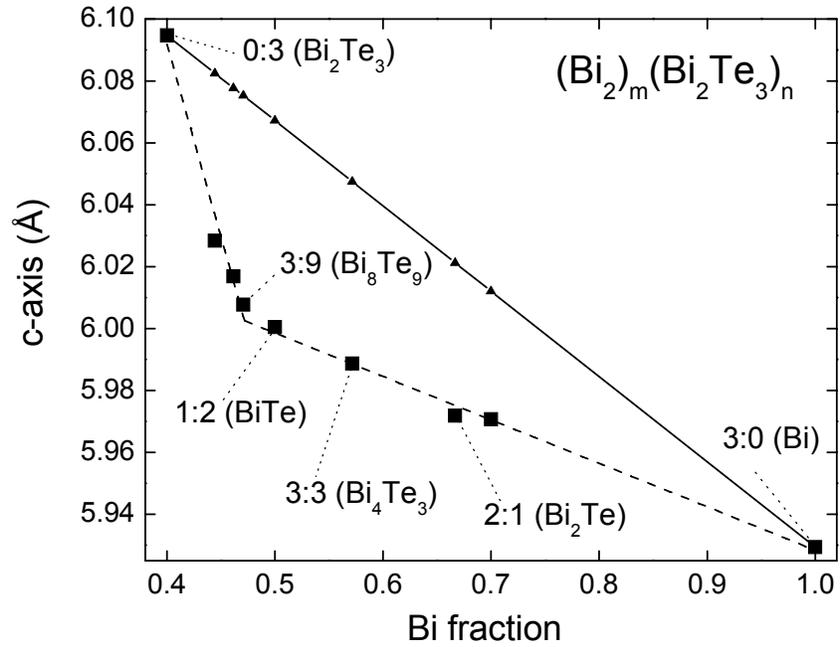

Fig. 4c

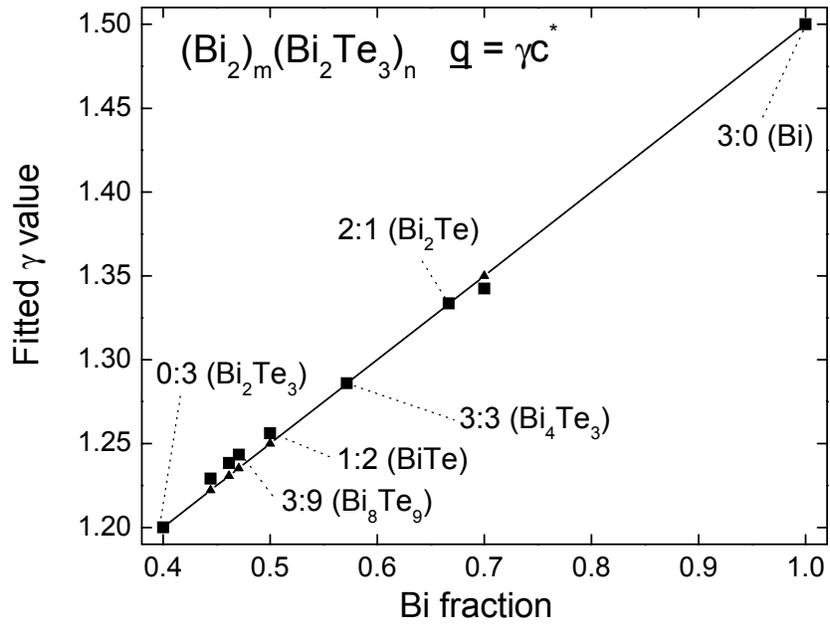

Fig 5.

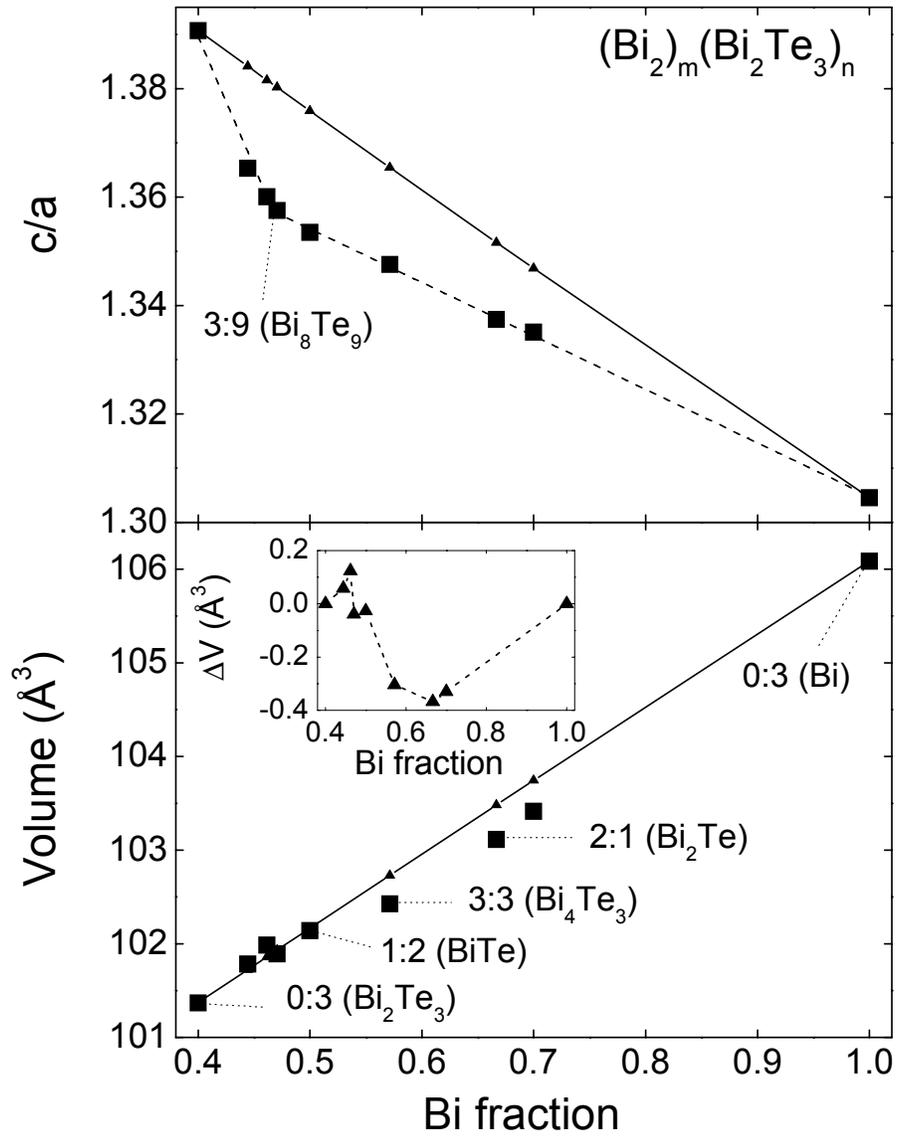

Fig. 6a

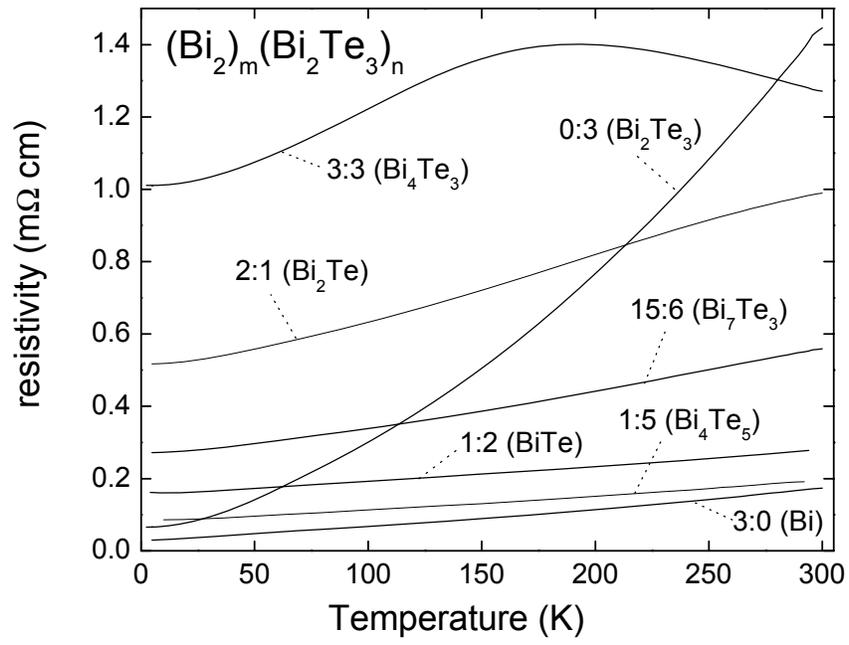

Fig. 6b

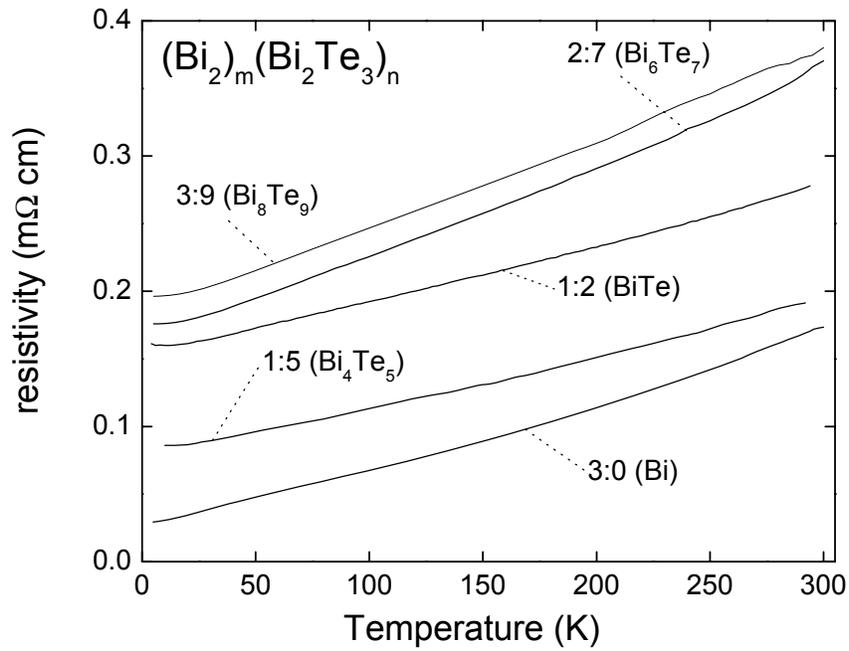

Fig. 7a

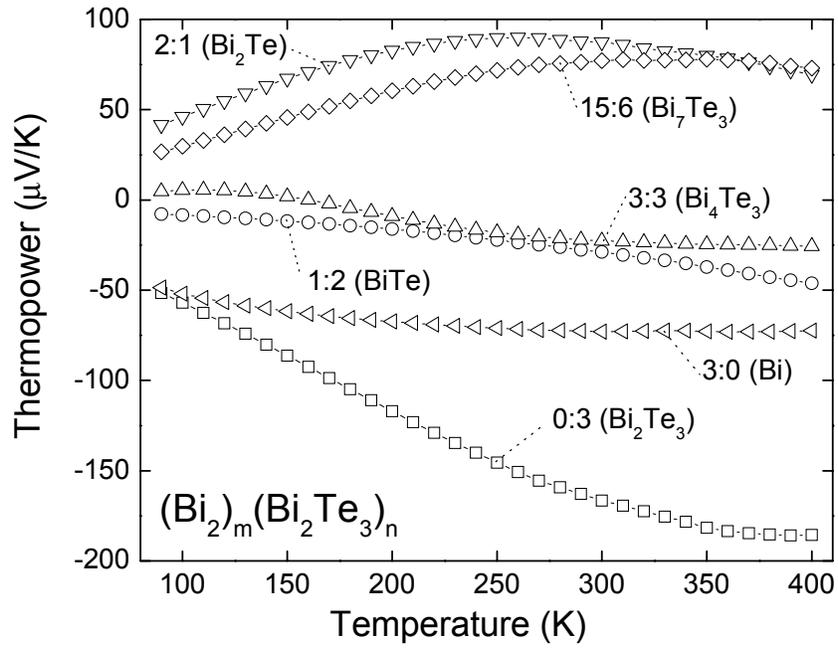

Fig. 7b

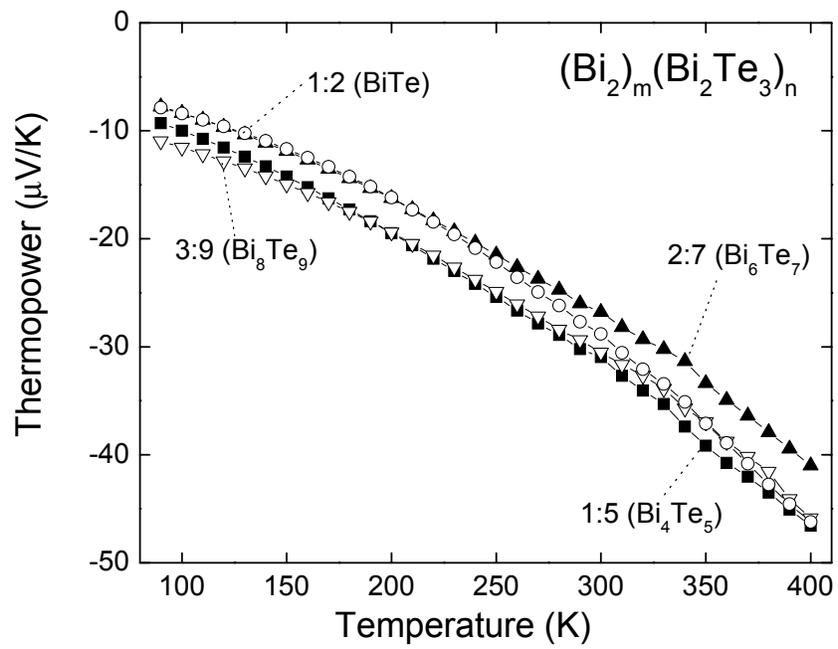

Fig. 8

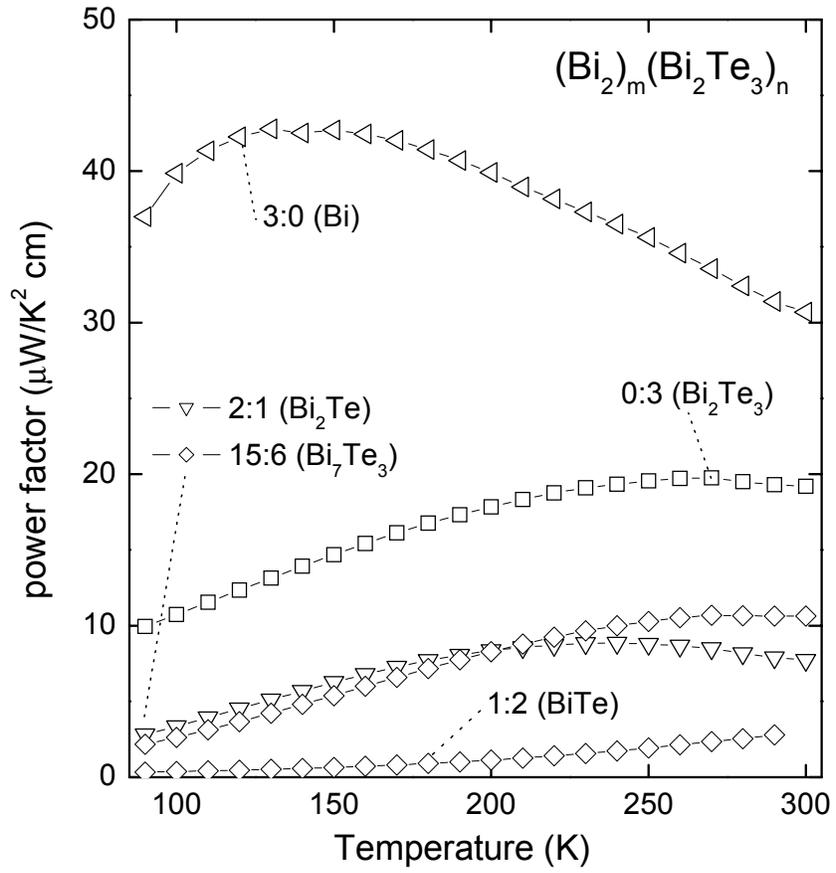

Fig. 9

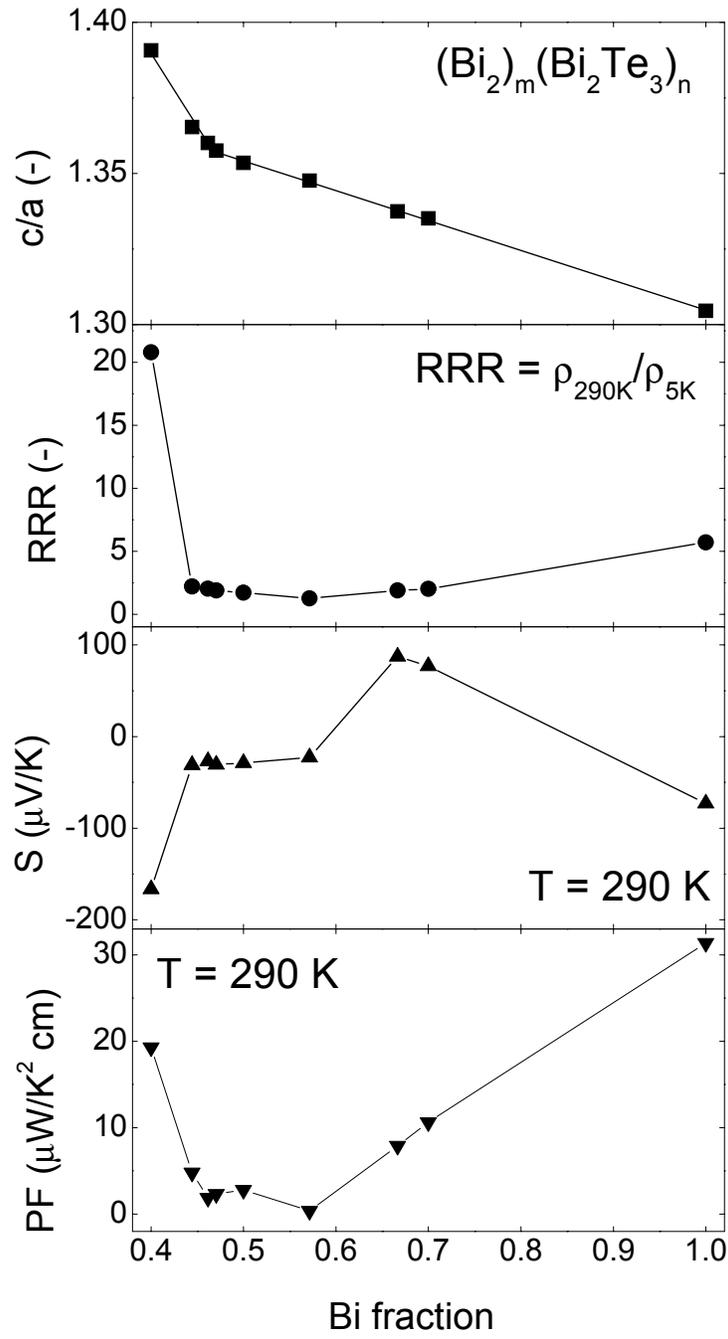

Fig. 10

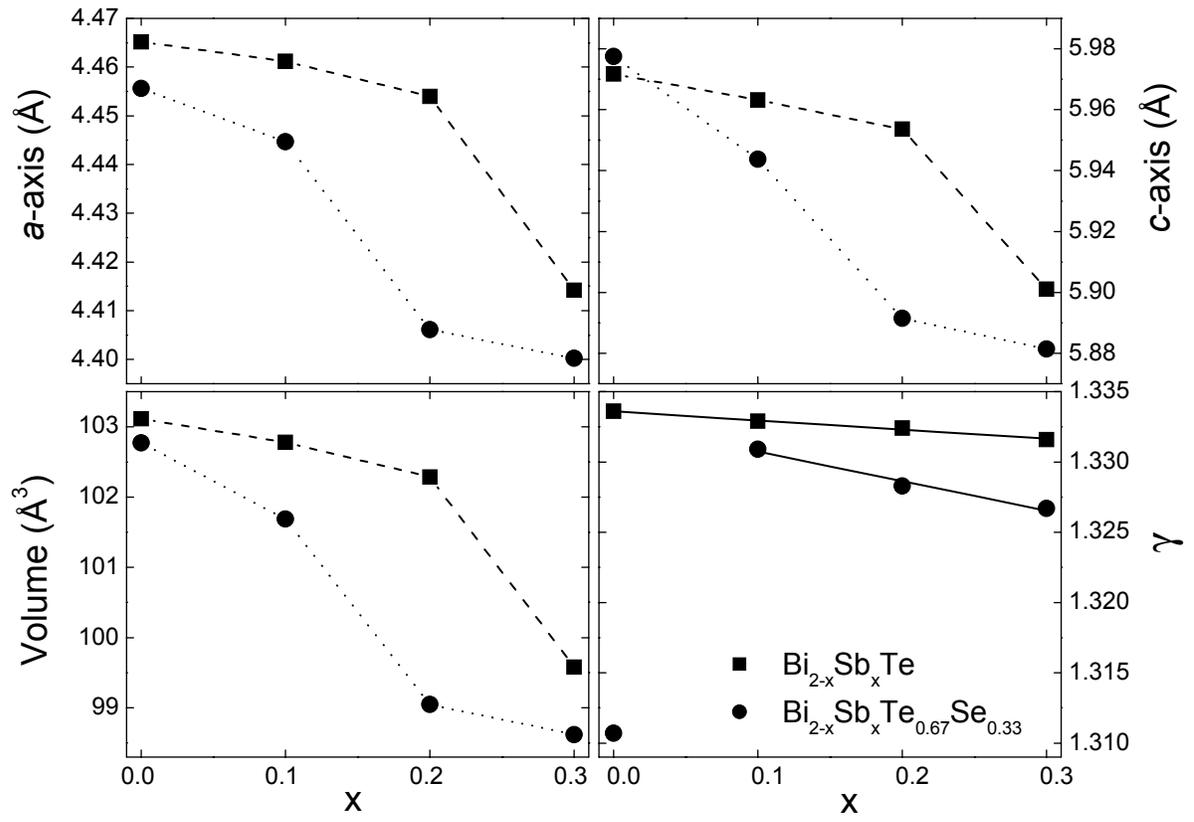

Fig. 11a

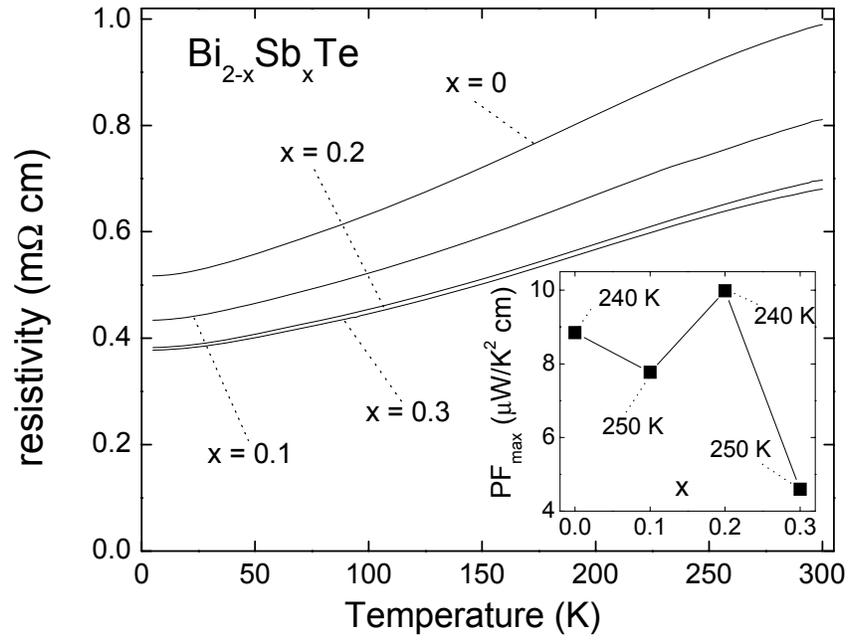

Fig. 11b

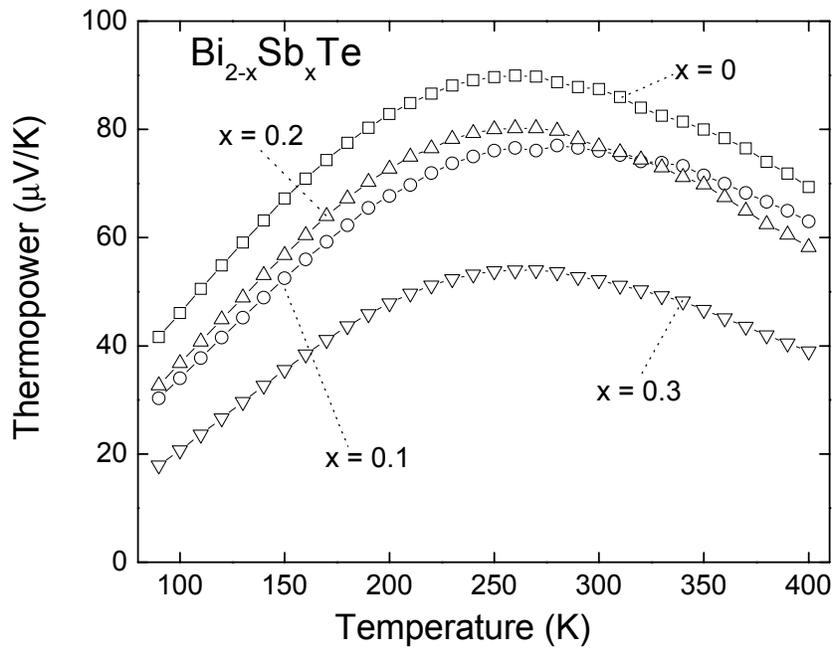

Fig. 12a

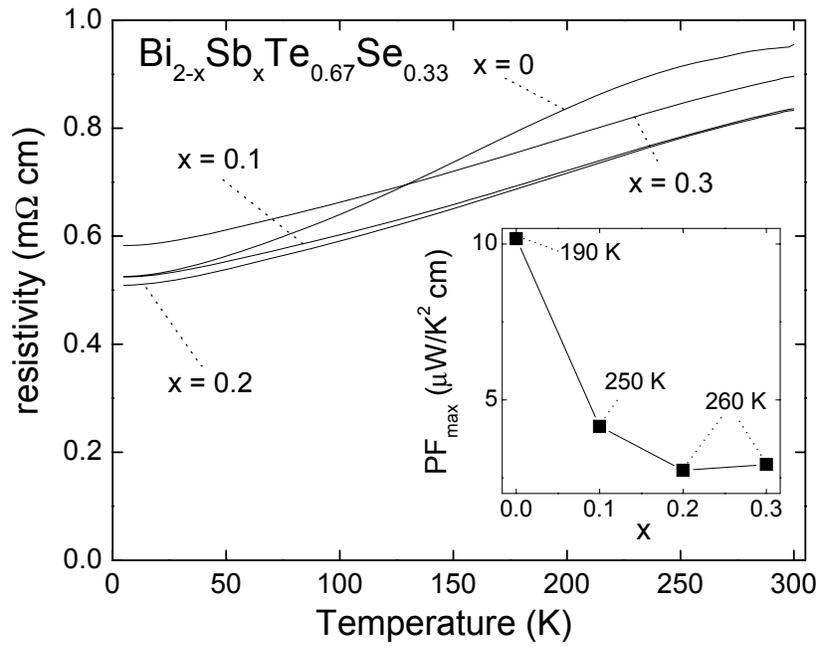

Fig. 12b

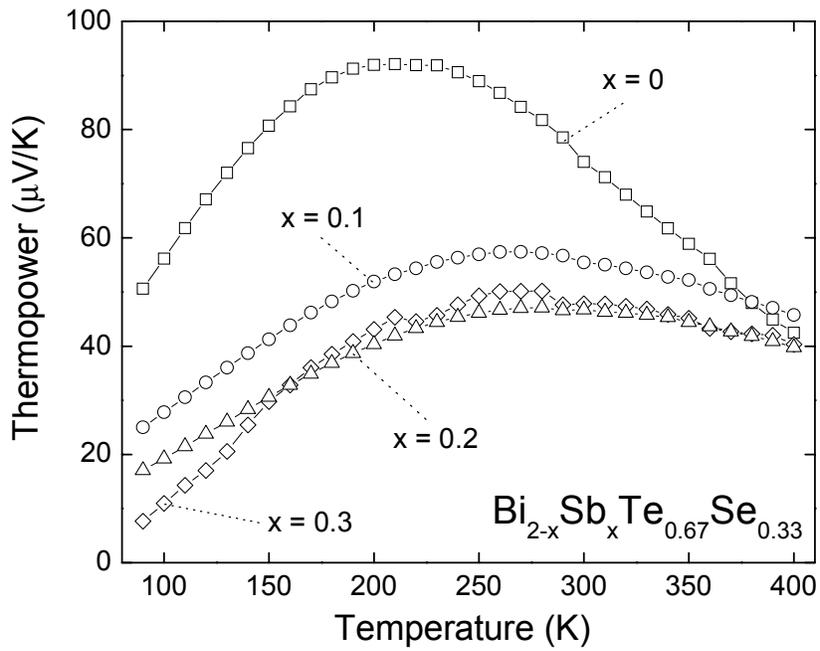